\documentclass[10pt,conference,twocolumn]{IEEEtran}
\usepackage{graphicx,amssymb,amsmath,color}
\usepackage{subfigure}
\usepackage{algorithm}
\usepackage{url}
\usepackage[noadjust]{cite}
\newtheorem{lemma}{Lemma}%
\definecolor{orange}{RGB}{255,107,0}
\definecolor{green} {RGB}{0,180,80}

\graphicspath{{Figs/}}
\title{A QoS-Aware Scheduling Algorithm for High-Speed Railway Communication System}
\author{Shengfeng Xu, Gang Zhu, Chao Shen, Bo Ai\\State Key Laboratory of Rail Traffic Control and
Safety\\Beijing Jiaotong University, Beijing, P. R. China
\\Email:$\{$11120177, gzhu, shenchao, boai$\}$@bjtu.edu.cn}
\begin{document}

\maketitle

\begin{abstract} 
With the rapid development of high-speed railway (HSR), how to provide the passengers with multimedia services has attracted increasing attention. A key issue is to develop an effective scheduling algorithm for multiple services with different quality of service (QoS) requirements. In this paper, we investigate the downlink service scheduling problem in HSR network taking account of end-to-end deadline constraints and successfully packet delivery ratio requirements.
Firstly, by exploiting the deterministic high-speed train trajectory, we present a time-distance mapping in order to obtain the highly dynamic link capacity effectively. Next, a novel service model is developed for deadline constrained services with delivery ratio requirements, which enables us to turn the delivery ratio requirement into a single queue stability problem. Based on the Lyapunov drift, the optimal scheduling problem is formulated and the corresponding scheduling service algorithm is proposed by stochastic network optimization approach.
Simulation results show that the proposed algorithm outperforms the conventional schemes in terms of QoS requirements.
\end{abstract}

\section{Introduction}
With the fast deployment of the high-speed railway (HSR) systems in the past several years, the demand for mobile communication on high-speed trains is increasingly growing. To fulfill the demand for streaming multimedia service on the train, the study of HSR communication is meaningful. For providing multiple services on the train, scheduling, as a critical parts of Radio Resource Management (RRM), is the related key issue to be solved \cite{Guan-2011}.

We consider a relay-assisted HSR network architecture in this work. The data packets are delivered via relay station (RS) to achieve a high data transmission rate instead of direct transmission \cite{Wang-2012}. For future HSR communication system, many kinds of real-time multimedia services should be provided on the train \cite{Barbu-2010}. These services often have different quality-of-service (QoS) requirements such as end-to-end deadline constraints and successfully packet delivery ratio requirements. Thus, for providing a large number of service deliveries, the resource contention among multiple services should be resolved and the efficient scheduling scheme should be proposed.

Many scheduling algorithms have been proposed for common communication systems to satisfy the QoS of multimedia services, e.g. \cite{Capozzi-2013} and \cite{Asadi-2013}. Also, many valuable efforts have been exerted for the design of scheduling algorithms with deadline constraints, e.g. \cite{Li-2012} and \cite{Lee-2013}. However, existing scheduling algorithms designed for common communication systems are not suitable to HSR communication system due to the highly dynamic link capacity. In addition, some unique characteristics in HSR communication system can greatly benefit the scheduling scheme design, e.g. cells distribute along the railway line and trains arrive on schedule.

Little existing papers study the scheduling algorithms in HSR networks.
Authors in \cite{Karimi-2012} proposed a scheduling and resource allocation mechanism in HSR networks with the cell array architecture, which maximizes the service rate by considering the channel variations and handover information.
\cite{Xu-2013} studied delay-aware fair resource allocation problem with heterogeneous packet arrivals and delay requirements for multiple services delivery in a relay-assisted HSR network.
Authors in \cite{Liang-2012} studied the optimal resource allocation problem in a cellular/infostation integrated HSR network, considering the intermittent network connectivity and multi-service demands.
However, the formulations and solutions cannot be directly applied to this paper because different QoS requirements for multiple services are considered, including end-to-end deadline constraints and successfully packet delivery ratio requirements.
Moreover, how to design an effective scheduling algorithm considering the unique characteristics in HSR communication systems is still an open problem.

In this paper, we investigate a multi-service scheduling problem in HSR communication systems, taking account of different packets arrival rates and QoS requirements.
Firstly, a time-distance mapping is set which exploits the deterministic high-speed train trajectory and the link capacity is obtained by considering time-varying path loss.
Next, we develop a novel service model for deadline constrained services with delivery ratio requirements, which enables us turn the delivery ratio requirement into a single queue stability problem.
And then the optimal scheduling problem is formulated based on the Lyapunov drift and the corresponding scheduling service algorithm is proposed by stochastic network optimization approach.
Simulation results show that compared with the traditional scheduling schemes, the proposed scheduling algorithm can effectively improve QoS requirements.

The remainder of the paper is organized as follows. Section II describes the system model. The problem formulation and transformation are provided in Section III. An effective scheduling algorithm is proposed in section IV. Some numerical results and discussions are shown in Section V. Finally, conclusions and future work are given in Section VI.
\section{System Model}
\begin{figure}[!h]
    \centering
    \includegraphics[scale = 0.61]{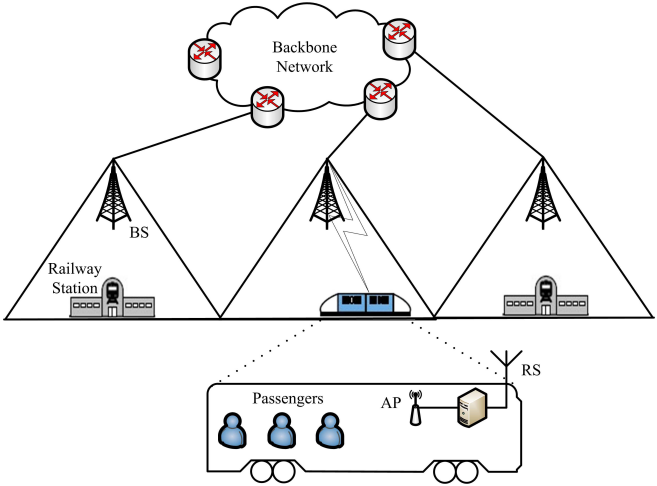}
    \caption{System Model}\label{fig:SystemModel-DC}
\end{figure}
A relay-assisted HSR network architecture is shown in Fig. \ref{fig:SystemModel-DC}. The cellular network deployed along the railway line can provide seamless coverage and data packets delivery. The basic stations (BSs) are connected to the backbone network via wireline links. The RS with powerful antennas is installed on the train to communicate with the BSs. The RS is further connected to an access point (AP) which can be accessed by the passengers based on wireless local area network (WLAN) technologies. Thus, with the help of RS, radio signals do not need to penetrate into the carriages so that radio signal penetration loss problem can be avoided. In addition, a packet can be successfully received by a passenger device if it is delivered to the RS due to the good channel in the carriages.

Compared to the common cellular networks, the deterministic train trajectory in HSR networks is a unique feature which represents the location of a train at a specific time. Since the train moves on a predetermined rail line and the velocity is relatively steady, the information of train trajectory can be obtained in advance with high accuracy so that the service packets can be delivered by the specific BS at certain time. Therefore, this paper focus on the downlink service scheduling problem regardless of which BS is used for service delivery.

In addition, for significantly simplifying the protocol design for HSR applications, erasure coding based service delivery is considered \cite{Liang-2012} and the advantage is that no recovery scheme is required for the transmission error or loss of specific packets due to highly dynamic wireless channel condition.

\subsection{Time-Distance Mapping}
Because of the characteristics of HSR communication systems, when a train starts at the original station, provided that the train is detected in the centre of the cell coverage and near to the BS, the channel condition from BS to RS is generally good with a highest value. As the train moves and time passes, the link quality degrades until to the edge of the cell, and then the link quality will start to improve as the vehicle moves away from the cell edge. The trains will experience the same link quality evolution while the evolution will be faster for trains traveling at a higher speed.

Consider a train travels from the origin station to a destination station within the time duration $[0,T_{O}]$ at a constant speed $v$. The distance which the train has traveled until time $t$ is $s(t) = vt$. We set $R$ be the cell radius, so the distance in a certain cell is $s_{1}(t) = s(t)~\text{mod}~2R$. Define a time-distance mapping function $d(t): [0,T_{O}] \rightarrow [d_{0},d_{\text{max}}]$, where $d_{\text{max}} = \sqrt{R^{2} + d_{0}^{2}}$ and $d_{0}$ is the distance between each BS and the rail line, which maps time $t$ to the corresponding distance from BS to RS. The mapping function $d(t)$ is given by
\begin{equation}\label{Time-Distance Mapping}
d(t) = \left\{\begin{array}{ccc}
\sqrt{s_{1}(t)^{2} + d_{0}^{2}},     &\text{if}~~0 \leq s_{1}(t) < R \\
\sqrt{(2R-s_{1}(t))^{2} + d_{0}^{2}},&\text{if}~~ R \leq s_{1}(t) < 2R
\end{array} \right.
\end{equation}
which is periodic function with the period of $T$, where $T = 2R/v$ is the time for trains spend on getting through one cell. As the speed $v$ increases, $T$ decreases and the number of data packets delivered to the train under a cell will decrease.

\subsection{Physical-layer Model}
For HSR communication networks, the average SNR of the received signal cannot remain at the same level, due to the fast-changing distance between the BS and the RS. As shown in \cite{Lin-2012}, the path loss $PL$ can be described as
\begin{equation}\label{path loss}
  PL(d)  = \left\{ \begin{array}{ll}
           44.2 + 21.5\log(d) + L & \textrm{$d < d_{bp}$},\\
           44.2 + 40\log(d/d_{bp}) + L_{bp} + L & \textrm{$d \geq d_{bp}$},
                   \end{array} \right.
\end{equation}
where $d$ is the distance from RS to BS. $L$ and $L_{bp}$ are constants in dB, $L = 20\log(f_{c}/(5\times 10^{9}))$ and $L_{bp} = 21.5\log(d_{bp})$. $d_{bp}$ is the break point of the path-loss curve, which is equal to $4h_{BS}h_{RS}f_{c}/c$, where $h_{BS}$ and $h_{RS}$ are the BS and RS antenna heights in meter compared to the ground, $f_{c}$ is the carrier frequency in Hz, and $c$ is velocity of light in vacuum.

Since the location information of a train at a specific time can be obtain with high accurate \cite{Liang-2012}, the path loss $\tilde{PL}(t)$ at time $t$ can be obtain based on (\ref{Time-Distance Mapping}) and (\ref{path loss}). Then the average received SNR can be denoted by
\begin{equation}\label{average received SNR}
         \overline{\gamma}(t) = P_{t} - \tilde{PL}(t) - N_{0}.
\end{equation}
where $P_{t}$ is the transmit power of the BS and $N_{0}$ is the noise power. The transmission rate can be expressed as follows:
\begin{equation}\label{transmission rate}
  R(t) = W\log_{2}(1+\overline{\gamma}(t))
\end{equation}
where $W$ is the system bandwidth. Since the packets have equal size of $G$ bits, the capacity $C$ of frame $k$ corresponding to time $t$ can be denoted as the maximum number of packets, i.e., $C[k] = \lfloor R(t)T_{F}/G \rfloor$, where $\lfloor x \rfloor = \max \{n\in \mathbb{Z} | n \leq x\}$ and $T_{F}$ is the frame length.

\subsection{Service Model}
The HSR network can support $S$ types of services and maintain one queue $Q_{s}$ for each service $s$. Let $\mathbf{A}[k] = (A_{1}[k],\ldots,A_{S}[k])$ represents the packet arrival vector at frame $k$ and $\mathbf{Q}[k] = (Q_{1}[k],\ldots,Q_{S}[k])$ represents the vector of current queue backlogs, where $Q_{s}[k]$ denotes the number of packets in $Q_{s}$ at the beginning of the frame $k$.

\emph{Assumption 1}: The packet arrival process for each service is assumed to be i.i.d. across frames. $\{ A_{s}[k] \}$ forms an ergodic Markov chain and Poisson distributed bursty traffic with average packet arrival rate $\lambda_{s} = \mathbb{E}[A_{s}[k]]$ for service $s$. The distribution of $A_{s}[k]$ for Poisson traffic can be given as
\begin{equation}\label{Poisson distribution}
  \mbox{Pr}[A_{s}[k]=i] = \exp(-\lambda_{s}) \frac{(\lambda_{s})^{i}}{i!},~i = 0,\ldots ,a_{s},
\end{equation}
where the truncated Poisson distribution with maximum number of incoming packets per frame, $a_{s}$ is found assuming $\mbox{Pr}[A_{s}[k]=a_{s}] \rightarrow 0$ and normalizing the distribution.

For capturing the demands of multimedia transmissions and real-time communication applications on the train, the QoS requirements of each service shall include a fixed end-to-end delay constraint and a delivery ratio requirement. On the one hand, each packet for service $s$ has a fixed end-to-end delay constraint of $m_{s}$ frames, after which the packet expires and will be dropped. On the other hand, service $s$ has a minimal delivery ratio requirement of $q_{s} \in (0,1)$, i.e., its loss probability upper bound is $p_{s} = 1-q_{s}$.

For feasibility of analysis, we impose the following necessary condition on the relationship between delivery ratio requirements of the services and average link capacity.

\emph{Assumption 2}: Given the delivery ratio requirements $q_{s}, s \in \mathcal{S}$, the arrival rate of the services satisfies
\begin{equation}\label{}
  \sum_{s=1}^{S} \lambda_{s}q_{s} \leq \frac{1}{K} \sum_{k=1}^{K} C[k]
\end{equation}
where $K = T_{O}/T_{F}$, indicating the total amount of frames.

Assumption 2 suggests that the average system capacity can support for $S$ types of services with minimal delivery ratio requirements.

\section{Problem Formulation and Transformation}
In this section, we first present the deadline constrained service formulation. Then the delivery ratio requirement is transformed into a single queue stability problem. Finally, the optimal scheduling problem is formulated based on the Lyapunov drift.
\subsection{Deadline Constrained Service Formulation}
We define $Q_{s}^{r}[k]$ to be the number of packets in $Q_{s}$ that have $r$ frames left before their expiry (those packets are called as packets having $r$ frames-to-go in the sequel) at the beginning of frame $k$. Apparently, we have
\begin{equation}\label{}
\sum_{r=1}^{m_{s}} Q_{s}^{r}[k] = Q_{s}[k]~~~~\forall s \in \mathcal{S}, \forall k
\end{equation}

In frame $k$, the packets delivery for service $s$ is denote by a vector $X_{s}[k] = \{X_{s}^{r}[k]\}_{r}$, where $X_{s}^{r}[k]$ is the number of packets being served that have $r$ frames-to-go at the beginning of frame $k$. We have the following capacity constraint for the each frame:
\begin{equation}\label{capacity constraint}
  \sum_{s=1}^{S}\sum_{r=1}^{m_{s}} X_{s}^{r}[k] \leq C[k]~~~~\forall k
\end{equation}
Also, to prevent the overutilization of the allocated resource, we need
\begin{equation}\label{}
  X_{s}^{r}[k] \leq Q_{s}^{r}[k]~~~~\forall k,r,s
\end{equation}

At the end of the frame $k$, we denote $D_{s}^{r}[k]$ to be the number of unserved packets of service $s$ which have $r$ frames-to-go at the beginning of frame $k$. It then follows from above that $D_{s}^{r}[k] = Q_{s}^{r}[k] - X_{s}^{r}[k]$. In particular, $D_{s}^{1}[k]$ represents the number of dropped packets of service $s$, which can be got by
\begin{equation}\label{dropped packets}
  D_{s}^{1}[k] = A_{s}[k-m_{s}+1] - \sum_{i=1}^{m_{s}} X_{s}^{m_{s}+1-i}[k-m_{s}+i]
\end{equation}
where the last term indicates the total numbers of packets allocated to $A_{s}[k-m_{s}+1]$.
Based on the above analysis, we have the following queue evolution equation for service $s$:
\begin{equation}\label{queue dynamics}
  Q_{s}^{r}[k+1] = \left\{
  \begin{array}{ll}
  A_{s}[k+1],       &  r = m_{s}\\
  D_{s}^{r+1}[k],   &  r = 1, \ldots, m_{s}-1
  \end{array}\right.
\end{equation}
Note that the above queue evolution is different from a standard one, since those unserved packets at the frame $k$ which have $r+1$ frames-to-go will have $r$ frames-to-go at the frame $k+1$.

\subsection{Delivery Ratio Requirements Transformation}
For better describing delivery ratio requirement, we introduce the deficit queue $Y_{s}$ for each service $s$.
The deficit queue $Y_{s}$ evolves according to the following equation:
\begin{equation}\label{deficit queue}
  Y_{s}[k+1] = (Y_{s}[k] - p_{s}\lambda_{s})^{+} + D_{s}^{1}[k]
\end{equation}
where $(x)^{+} = x$ if $x > 0$, and $0$ otherwise, and $Y_{s}[0] = 0$ for all services.
Thus, the process $Y_{s}[k]$ for service $s$ acts as a single-server queue with constant server rate $p_{s}\lambda_{s}$ given by the maximal packet loss probability and average packet arrival rate, with arrival rate given by the number of dropped packets $D_{s}^{1}[k]$ at the frame $k$. In addition, the deficit queues can be regarded as counters rather than actual queues that store the expired packets.

\begin{lemma}
If the deficit queue is rate stable, i.e., satisfies
\begin{equation}\label{}
  \lim\limits_{k \rightarrow \infty} \frac{Y_{s}[k]}{k} = 0
\end{equation}
then
\begin{equation}\label{minimal delivery ratio constraint}
  \overline{D_{s}^{1}} = \lim\limits_{k \rightarrow \infty} \frac{1}{k} \sum_{l=0}^{k-1} D_{s}^{1}[l] \leq p_{s}\lambda_{s}.
\end{equation}
\end{lemma}

The intuition behind Lemma 1 is given by the following observation: For each service $s \in \mathcal{S}$, if the deficit queue $Y_{s}[k]$ is stable, then the corresponding minimal delivery ratio requirement can be satisfied. This observation holds because if the excess backlog in deficit queue is stabilized, it must be the case that the time average arrival rate (corresponding to $\overline{D_{s}^{1}}$) is less than or equal to the service rate (corresponding to $p_{s}\lambda_{s}$).
The proof of Lemma 1 is provided in Appendix A. Based on the Lemma 1, the minimal delivery ratio constraint (\ref{minimal delivery ratio constraint}) can be turn into a single queue stability problem.

\subsection{Scheduling Problem Formulation}
The queue evolution given in (\ref{queue dynamics}) implies that the arrivals $A_{s}[k]$ in frame $k$ are only to be served by $\mathbf{X}_{s}[k, m_{s}] \triangleq \{X_{s}^{m_{s}}[k], X_{s}^{m_{s}-1}[k+1], \ldots, X_{s}^{1}[k+m_{s}-1]\}$ before their expiry. Based on this observation, an effective scheduling algorithm should decide on $\mathbf{X}_{s}[k, m_{s}]$ for all services in each frame based on current and past information. The specific scheduling problem formulation is described as follows.

First, we define $\boldsymbol{Y}[k]$ as the combined vector of all deficit queues:
$\boldsymbol{Y}[k] \triangleq \{Y_{s}[k+m_{s}-1]\}_{s}$, which is the observed network event at frame $k$.
Next, the Lyapunov function is given as follows:
\begin{equation}\label{Lyapunov function}
  L(\boldsymbol{Y}[k]) \triangleq \frac{1}{2} \sum_{s\in \mathcal{S}} (Y_{s}[k+m_{s}-1])^{2}.
\end{equation}
Finally, $\Delta(\boldsymbol{Y}[k])$ is defined as the one-frame conditional Lyapunov drift at frame $k$:
\begin{equation}\label{Lyapunov drift}
  \Delta(\boldsymbol{Y}[k]) = \mathbb{E}[L(\boldsymbol{Y}[k+1]) - L(\boldsymbol{Y}[k])|\boldsymbol{Y}[k]]
\end{equation}

The objective of the downlink scheduling problem is to minimize the drift by designing a scheduling algorithm over a trip of the train. The optimal problem is formulated as
\begin{subequations}\label{optimal problem}
\begin{align}
   \mbox{Minimize}: &~~\Delta(\boldsymbol{Y}[k]) \label{maximize}\\
   \mbox{Subject to}:            &~~\sum_{s=1}^{S}\sum_{r=1}^{m_{s}} X_{s}^{r}[k] \leq C[k]~~~~\forall k \label{st-1}\\
                                 &~~X_{s}^{r}[k] \leq Q_{s}^{r}[k]~~~~\forall k,r,s\label{power constraint}
\end{align}
\end{subequations}
This optimal problem is a stochastic optimization problem and can be solved by using a stochastic network optimization approach\cite{Neely-2010}, which will be discussed in the following section.

\section{Deadline Constrained Scheduling Algorithm}
In this section, we develop a scheduling framework with deficit queue stability based on a stochastic network optimization approach and propose a deadline constrained scheduling algorithm.
\begin{lemma}
For any initial deficit queue values $\boldsymbol{Y}[k]$, one-frame conditional Lyapunov drift under any scheduling decisions satisfies:
\begin{align}\label{Lemma 2-1}
  \Delta(\boldsymbol{Y}[k]) \leq B - \mathbb{E}\left[ \sum_{s\in \mathcal{S}} Y_{s}[k+m_{s}-1]p_{s}\lambda_{s}  |\boldsymbol{Y}[k] \right]& \nonumber\\
  + \mathbb{E}\left[ \sum_{s\in \mathcal{S}} Y_{s}[k+m_{s}-1]D_{s}^{1}[k+m_{s}-1] |\boldsymbol{Y}[k] \right]&
\end{align}
where $B$ is a finite constant defined:
\begin{equation}\label{Lemma 2-2}
  B \triangleq \frac{1}{2} \sum_{s\in \mathcal{S}} \left[ (p_{s}\lambda_{s})^{2} + (a_{s})^{2} \right]
\end{equation}
where we recall $a_{s}$ is the bound in Assumption 1.
\end{lemma}

The details of the proof of Lemma 2 are provided in Appendix B.
Based on the Lemma 2, the design principle behind our scheduling algorithm is apparent: Given at frame $k$, to observe the current deficit queue backlogs, the algorithm make a scheduling decision to minimize the right-hand-side of the drift bound (\ref{Lemma 2-1}). Note that the scheduling decision $\mathbf{X}_{s}[k, m_{s}]$ for all services are independent of the second term on the right-hand-side of the equation (\ref{Lemma 2-1}) and only affects the final term on the right-hand-side of the equation (\ref{Lemma 2-1}). Thus, we seek to design an algorithm that minimizes the following expression:
\begin{equation}\label{Lemma 2-3}
  \mathbb{E}\left[ \sum_{s\in \mathcal{S}} Y_{s}[k+m_{s}-1]D_{s}^{1}[k+m_{s}-1] |\boldsymbol{Y}[k] \right]
\end{equation}
The above conditional expectation is with respect to the randomly arrival service rate $\mathbf{A}(t)$ and the possibly scheduling decision. Based on further investigation, the concept of opportunistically minimizing an expectation in \cite{Neely-2010} can be applied to the above expression, which is minimized by observing the current deficit queues $\boldsymbol{Y}[k]$ and service arrival states $\mathbf{A}(t)$ and choosing the scheduling decision $\mathbf{X}_{s}[k, m_{s}]$ for all services to minimize:
\begin{equation}\label{Lemma 2-4}
   \sum_{s\in \mathcal{S}} Y_{s}[k+m_{s}-1]D_{s}^{1}[k+m_{s}-1],
\end{equation}
which can be solved by the weight-based algorithm \cite{Neely-2010}. Specifically, the minimal of (\ref{Lemma 2-4}) can be achieved by obtaining the smallest possible value of each $D_{s}^{1}[k+m_{s}-1]$ in the order of decreasing value of $Y_{s}[k+m_{s}-1]$. Based on the above framework, we propose a deadline constrained scheduling algorithm (DCSA) as shown in Algorithm 1.

\begin{algorithm}\caption{DCSA}\label{a-1}
0:~Initialize $\mathbf{X}_{s}[k, m_{s}] = \mathbf{0}$, for $s \in \mathcal{S}$;\\
1:~$Y_{s}[k+m_{s}-1]$ calculation, for $s \in \mathcal{S}$;\\
2:~Sort services in descending of $Y_{s}[k+m_{s}-1]$ and obtain
$~~~~~$the ordered set $\{s_{1},s_{2},\ldots,s_{|\mathcal{S}|}\}$;\\
3:~~\textbf{for} $n = 1$ to $|\mathcal{S}|$ \textbf{do}\\
4:~~~~\textbf{for} $i = 0$ to $m_{s_{n}}-1$ \textbf{do}\\
5:~~~~~~$Z[k+i] = \sum_{\tilde{s}=1}^{S}\sum_{r=1}^{m_{\tilde{s}}-i-1} X_{\tilde{s}}^{r}[k+i]$;\\
6:~~~~~~$X_{s_{n}}^{m_{s_{n}}-i}[k+i] = \min\Big(A_{s_{n}}[k]- \sum_{j=0}^{i-1} X_{s_{n}}^{m_{s_{n}}-j}[k+j],$\\
$~~~~~~~~~~~~~~~~~~~C[k+i] - Z[k+i]- \sum_{h=1}^{n-1} X_{s_{h}}^{m_{s_{h}}-i}[k+i]\Big)$;\\
7:~~~~\textbf{end for}\\
8:~~\textbf{end for}\\
9:~Update the deficit queue $Y_{s}$ by (\ref{deficit queue});
\end{algorithm}

As the train moves from the origin station to the destination terminal, the Algorithm 1 allocates the network
resources to multiple services frame-by-frame. Specifically, In each frame $k$, the scheduler observes $\boldsymbol{Y}[k]$ and the arrival state $\mathbf{A}[k]$ as well as the remaining link capacity, and makes a scheduling action $\mathbf{X}_{s}[k, m_{s}]$. The link capacity is allocated to the multiple services in descending order of $Y_{s}[k+m_{s}-1]$, until it is fully utilized or all arrival services are scheduled. In step 5, $Z[k+i]$ represents the total resource at frame $k+i$ allocated to the packets that arrived before the frame $k$. The minimum operation in step 6 is performed over two terms corresponding to the unserved packets and the remaining link capacity, respectively. For the first term, the summation $\sum_{j=0}^{i-1} X_{s_{n}}^{m_{s_{n}}-j}[k+j]$ is subtracted since these packets are scheduled before the frame $k+i$; and for the second term, the summation $Z[k+i]+ \sum_{h=1}^{n-1} X_{s_{h}}^{m_{s_{h}}-i}[k+i]$ is subtracted since the corresponding capacity has been used for the previous services and high-priority services at frame $k+i$. At the end of frame $k$, the deficit queue $Y_{s}$ is updated in step 9 by (\ref{deficit queue}). Note that the complexity of Algorithm 1 is low with $O(\sum_{s=1}^{S}m_{s})$.

\section{Simulation Results and Discussions}
\subsection{Simulation Setup}
For the purpose of comparison, we evaluate two related scheduling schemes as reference benchmarks. The first is the traditional round-robin (RR) scheme which schedules services in a predetermined order. At frame $k$, the earliest packets within the deadline of the ($k$ mod $S+1$)-th service is chosen for transmitting. The second is the Earliest Deadline First (EDF) scheme \cite{Xu-1990}, which always servers the most urgent packet without a predetermined order.
We summarize the system parameters in Table I. A single simulation runs the algorithms when the train moves through a cell (30000 frames) and we plot the first 1000 frames for clarity.

\begin{table}[!htb]
\renewcommand{\arraystretch}{1.3}
\caption{Parameters in simulation \label{Table1}}
\centering
\begin{tabular}{cc|cc}
\hline Parameter & Value & Parameter & Value\\
\hline $R$&1.5 km&$v$& 100 m/s\\
\hline $T_{F}$&$1$ ms&$W$&10 MHz\\
\hline $G$&500 bits&$P_{t}/N_{0}$&115 dB\\
\hline $h_{RS}$&5 m&$h_{BS}$&50 m\\
\hline $d_{0}$&30 m&$f_{c}$&2.4 GHz\\
\hline
\end{tabular}
\end{table}

\subsection{Simulation Results}
\begin{figure}
  \centering
  \includegraphics[scale=0.68]{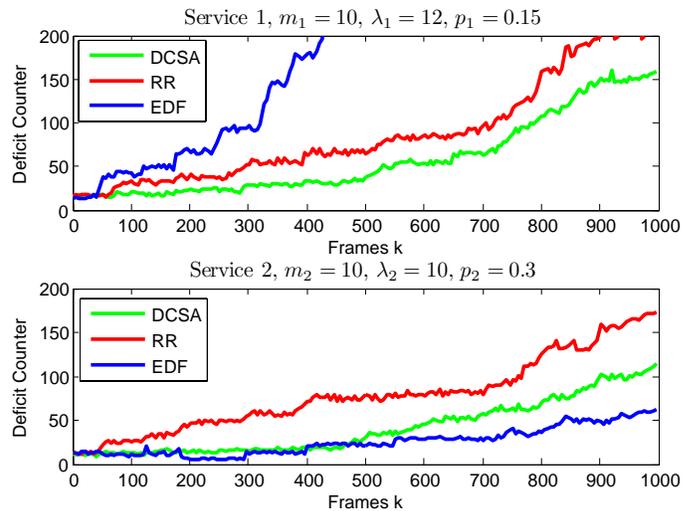}\\
  \caption{Evolution of the deficit counter $Y_{s}[k]$ for $S=2$.}\label{1}
\end{figure}

For comparing the performance of packet loss probability for these three schemes, we set simulations by observing the evolution of the deficit counter values, because a smaller deficit counter value implies lower packet loss probability. Fig. \ref{1} describes the evolution of deficit counters for two services with different arrival rates and delivery ratio requirements.
Service 1 has a higher arrival rate as well as a higher delivery ratio requirement compared to service 2.
By simply serving the packet which is closest to its expiry, the EDF scheme is ignorant to the difference in the delivery ratio requirements, resulting in the insufficient allocation to service 1, whose deficit counter increases rapidly.
The RR scheme provides the equal scheduling chance to these two services regardless of the different QoS requirements.
In contrast, the DCSA algorithm considers the long-term delivery ratio requirements by using the deficit queues and tries to balance them for these two services.
The above analysis illustrates the advantages of the proposed DCSA algorithm, which could effectively reduce the packet loss probability as well as balance the different QoS requirements in the HSR networks.
In addition, the values of deficit counter increase at the first 1000 frames, because the train is moving from center to edge of one cell and the link capacity degrades.

\begin{figure}
  \centering
  \includegraphics[scale=0.68]{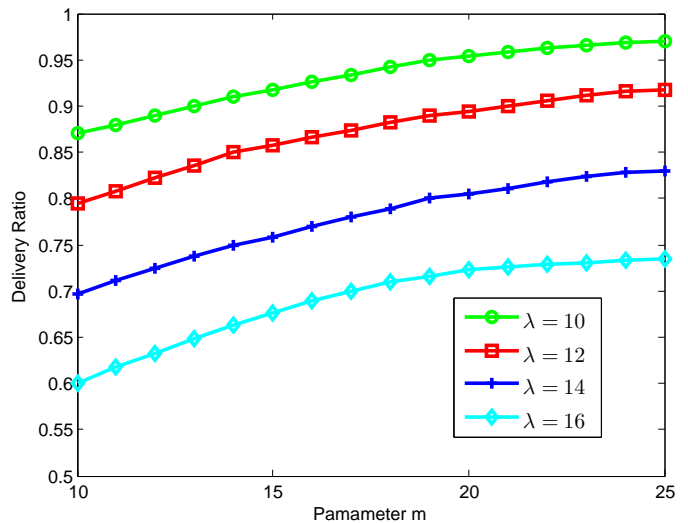}\\
  \caption{Delivery ratio with different deadline constraints $m$}\label{2}
\end{figure}

Fig. \ref{2} illustrates the achieved delivery ratio for different values of deadline constraints and arrival rates considering one service situation. It can be observed that the high arrival rate will result in the degradation of the delivery ratio performance. This can be explained as follows: Since the link capacity is fixed and finite in general, when the number of arrival packets per frame increases, there are more data packets to be transmitted before the deadline, then more packets will be dropped. Furthermore, we can see that, as the parameter $m$ increases, the delivery ratio grows. This is because a larger $m$ will leave more time freedom in scheduling the packets and more link capacity can be utilized before one's deadline.

\section{Conclusions}
Providing passengers with multimedia services is one of the most important applications in HSR communication system. In this paper, a novel service model has been developed for deadline constrained services with delivery ratio requirements in HSR networks. The novelty of the proposed scheduling algorithm lies in stochastic network optimization approach, and the application of ideas such as the minimal delivery ratio constraint transformation and opportunistically minimizing an expectation.
Simulation results were present to demonstrate the performance of the proposed scheduling algorithm in terms of packet loss probability and deadline assurance. Since this paper just considered the link from BS to RS, we will study the deadline constrained scheduling problem for the relay-assisted HSR network with two-hop link in our future work.

\begin{appendices}
\section{}
\emph{Proof of Lemma 1:} For any frame $l$, we have by (\ref{deficit queue}):
\begin{displaymath}
Y_{s}[l+1] = (Y_{s}[l] - p_{s}\lambda_{s})^{+} + D_{s}^{1}[l]
\end{displaymath}
Thus, $Y_{s}[l+1] - Y_{s}[l] \geq D_{s}^{1}[l] - p_{s}\lambda_{s}$ for all $l$.
Summing the above over $l \in \{0,\ldots, k-1\}$ yields $Y_{s}[k] - Y_{s}[0] \geq \sum_{l=0}^{k-1} (D_{s}^{1}[l] - p_{s}\lambda_{s})$. Dividing by $k$ and taking limit as $k \rightarrow \infty$, we can get
\begin{equation}\label{}
  \lim_{k \rightarrow \infty} \frac{Y_{s}[k]}{k} \geq \lim_{k \rightarrow \infty} \frac{1}{k}\sum_{l=0}^{k-1} (D_{s}^{1}[l]) - p_{s}\lambda_{s} = \overline{D_{s}^{1}} - p_{s}\lambda_{s}.
\end{equation}
Thus, if $\lim\limits_{k \rightarrow \infty} Y_{s}[k]/k = 0$, then $\overline{D_{s}^{1}} \leq p_{s}\lambda_{s}$ for $s \in \mathcal{S}.$~~~~~~~~$\blacksquare$

\section{}
\emph{Proof of Lemma 2:} Recall the evolution equation (\ref{deficit queue}) for each deficit queue $Y_{s}[k]$ at frame $k$, and by squaring this equation, we obtain
\begin{align}\label{squaring evolution equation}
  &(Y_{s}[k+1])^{2} \nonumber\\
  & \leq (Y_{s}[k] - p_{s}\lambda_{s})^{2} + (D_{s}^{1}[k])^{2} +2D_{s}^{1}[k](Y_{s}[k] - p_{s}\lambda_{s})^{+} \nonumber\\
  & \leq (Y_{s}[k])^{2} + (p_{s}\lambda_{s})^{2} + (D_{s}^{1}[k])^{2} +2Y_{s}[k](D_{s}^{1}[k] - p_{s}\lambda_{s})
\end{align}
where we have used the fact that for any $x,y\geq 0$, we have $(\max[x,0])^{2} \leq x^{2}$ in the first inequality and $\max[x-y,0] \leq x$ in the second inequality.
Based on the equation (\ref{Lyapunov drift}) and (\ref{squaring evolution equation}), we have
\begin{align}\label{}
  &\Delta(\boldsymbol{Y}[k]) \nonumber\\
  &= \mathbb{E}\left[\frac{1}{2} \sum_{s\in \mathcal{S}} \left[(Y_{s}[k+m_{s}])^{2} - (Y_{s}[k+m_{s}-1])^{2}\right]|\boldsymbol{Y}[k]\right]\nonumber\\
  &\leq \mathbb{E}\Bigg[\frac{1}{2} \sum_{s\in \mathcal{S}} \Big[ (p_{s}\lambda_{s})^{2} + (D_{s}^{1}[k+m_{s}-1])^{2} \nonumber\\
  & ~~~+2Y_{s}[k+m_{s}-1](D_{s}^{1}[k+m_{s}-1] - p_{s}\lambda_{s}) \Big] |\boldsymbol{Y}[k] \Bigg]\nonumber\\
  &\leq B + \mathbb{E}\left[ \sum_{s\in \mathcal{S}} Y_{s}[k+m_{s}-1](D_{s}^{1}[k+m_{s}-1] - p_{s}\lambda_{s}) |\boldsymbol{Y}[k] \right]
\end{align}
where the last inequality can be obtained as follows. For all frame $k$, all possible $\boldsymbol{Y}[k]$, and all possible scheduling decisions $\mathbf{X}_{s}[k, m_{s}]$ that can be taken, we have
\begin{align}\label{}
  &\mathbb{E}\left[\frac{1}{2} \sum_{s\in \mathcal{S}} \left[ (p_{s}\lambda_{s})^{2} + (D_{s}^{1}[k+m_{s}-1])^{2} \right] |\boldsymbol{Y}[k] \right] \nonumber\\
  \leq~&\mathbb{E}\left[\frac{1}{2} \sum_{s\in \mathcal{S}} \left[ (p_{s}\lambda_{s})^{2} + (A_{s}[k])^{2} \right] |\boldsymbol{Y}[k] \right] \nonumber\\
  \leq~&\frac{1}{2} \sum_{s\in \mathcal{S}} \left[ (p_{s}\lambda_{s})^{2} + (a_{s})^{2} \right] = B
\end{align}
where the first inequality holds due to $D_{s}^{1}[k+m-1] \leq A_{s}[k]$ based on the equation (\ref{dropped packets}), and the second inequality holds due to $A_{s}[k] \leq a_{s}$ in Assumption 2.~~~~~~~~~~~~~~~~~~~~~~~~~~~~~~~~~~~~~$\blacksquare$
\end{appendices}

\section*{Acknowledgment}
This work was supported by the Fundamental Research Funds for the Central Universities under Grant No. 2014YJS026 and 2013JBM147, the Key Project of State Key Lab of Rail Traffic Control and Safety under Grant No. RCS2012ZZ004, and the Natural Science Foundation of China under Grant No. U1334202.

\bibliographystyle{IEEEtran}
\bibliography{Ref_DCS}
\end{document}